\documentclass[10pt, conference, letterpaper]{IEEEtran}
\IEEEoverridecommandlockouts

\usepackage{amsmath,amssymb,amsfonts}

\usepackage{algorithmic}
\usepackage[linesnumbered,ruled,vlined, noend]{algorithm2e}
\usepackage{algorithmic}
\usepackage{textcomp}
\usepackage[T1]{fontenc}
\usepackage{cite}
\usepackage[hyphens]{url}
\usepackage{subcaption}
\usepackage{pifont}
\usepackage{tabularx}

\usepackage{enumitem}
\usepackage{graphicx}
\usepackage{xcolor}
\usepackage[cmintegrals]{newtxmath}
\usepackage{bm}
\usepackage{booktabs}
\usepackage{multirow}
\usepackage{footnote}
\usepackage{balance}
\usepackage{tikz}
\usepackage{cleveref}
\usepackage[referable]{threeparttablex}
\crefformat{section}{\S#2#1#3}
\crefformat{subsection}{\S#2#1#3}
\crefformat{subsubsection}{\S#2#1#3}

\SetKwInput{KwData}{Input}
\SetKwInput{KwResult}{Output}

\def\BibTeX{{\rm B\kern-.05em{\sc i\kern-.025em b}\kern-.08em
    T\kern-.1667em\lower.7ex\hbox{E}\kern-.125emX}}

\newcommand*\circled[1]{\tikz[baseline=(char.base)]{
            \node[shape=circle,draw,inner sep=0.4pt] (char) {#1};}}

\begin{document}

\title{OmniSense: Towards Edge-Assisted Online Analytics for 360-Degree Videos \\
\thanks{
This work is supported by a Natural Sciences and Engineering Research Council of Canada (NSERC) Discovery Grant, a Canada Foundation for Innovation (CFI) John R. Evans Leaders Fund (JELF, \#$40215$), and a British Columbia Knowledge Development Fund (BCKDF). Fangxin Wang's work is supported by the National Natural Science Foundation of China with Grant No. $62102342$ and The Major Key Project of PCL Department of Broadband Communication. Fangxin Wang and Jiangchuan Liu are the corresponding authors.}}

\author{
    \IEEEauthorblockN{
        Miao Zhang\IEEEauthorrefmark{1}, 
        Yifei Zhu\IEEEauthorrefmark{2},
        Linfeng Shen\IEEEauthorrefmark{1},
        Fangxin Wang\IEEEauthorrefmark{3}\IEEEauthorrefmark{4}, 
        Jiangchuan Liu\IEEEauthorrefmark{1}
    }
    \IEEEauthorblockA{
        \IEEEauthorrefmark{1} School of Computing Science, Simon Fraser University, Canada \\
        \IEEEauthorrefmark{2} Cooperative Medianet Innovation Center (CMIC), Shanghai Jiao Tong University, China \\
        \IEEEauthorrefmark{3} SSE and FNii, The Chinese University of Hong Kong, Shenzhen, China\\
        \IEEEauthorrefmark{4} Peng Cheng Laboratory, Shenzhen, China \\
        mza94@sfu.ca, yifei.zhu@sjtu.edu.cn, linfeng\_shen@sfu.ca, wangfangxin@cuhk.edu.cn, jcliu@cs.sfu.ca
    }
}
\maketitle

\begin{abstract}
With the reduced hardware costs of omnidirectional cameras and the proliferation of various extended reality applications, more and more $360^\circ$ videos are being captured. To fully unleash their potential, advanced video analytics is expected to extract actionable insights and situational knowledge without blind spots from the videos. In this paper, we present OmniSense, a novel edge-assisted framework for online immersive video analytics. OmniSense achieves both low latency and high accuracy, combating the significant computation and network resource challenges of analyzing $360^\circ$ videos. Motivated by our measurement insights into $360^\circ$ videos, OmniSense introduces a lightweight spherical region of interest (SRoI) prediction algorithm to prune redundant information in $360^\circ$ frames. Incorporating the video content and network dynamics, it then smartly scales vision models to analyze the predicted SRoIs with optimized resource utilization. We implement a prototype of OmniSense with commodity devices and evaluate it on diverse real-world collected $360^\circ$ videos. Extensive evaluation results show that compared to resource-agnostic baselines, it improves the accuracy by $19.8\%$ -- $114.6\%$ with similar end-to-end latencies. Meanwhile, it hits $2.0\times$ -- $2.4\times$ speedups while keeping the accuracy on par with the highest accuracy of baselines.
\end{abstract}


\section{Introduction}
Recent years have seen an increasing number of affordable omnidirectional cameras being released, e.g., Insta$360$ ONE X$2$ \cite{insta360-one} and GoPro MAX \cite{gopro-max}. Unlike conventional cameras that record videos only capturing a narrow field of view (FoV), omnidirectional cameras record $360^\circ$ videos \cite{qian2018flare} covering an omnidirectional FoV without blind spots. Although such $360^\circ$ videos are popular for providing immersive experiences for human viewers \cite{qian2018flare, chen2021popularity}, their full potential has yet to be reached. As true recordings of the physical world, the $360^\circ$ videos can further help humans, robots, and devices understand and interact with their surroundings if the video content can be analyzed automatically.

Inspired by the success of video analytics systems for regular videos \cite{zhang2017live, jiang2018chameleon, ran2018deepdecision}, we believe \emph{immersive video analytics}, which applies vision algorithms to $360^\circ$ video content for automated knowledge extraction, will become the key to unlocking the full potential of $360^\circ$ videos. In years to come, omnidirectional cameras combined with immersive video analytics will build critical immersive visual sensing interfaces for a variety of eXtended Reality (XR) applications. For instance, a self-driving car can be fully aware of its surroundings by analyzing the videos captured by a roof-mounted omnidirectional camera so as to navigate safely. Police officers on patrol can learn the big picture of activities taking place in large public spaces through the video analytics results of body-worn omnidirectional cameras to ensure that no emergencies go unnoticed. 

While promising, achieving low-latency and high-accuracy immersive video analytics faces severe computation and network resource challenges. Contemporary video analytics systems boost accuracy with deep neural networks (DNNs), which are known to be resource-intensive \cite{zhang2017live}. Typical mobile devices are not even equipped with sufficient computation resources to support analysis for regular videos \cite{li2020reducto, zhang2021elf}, let alone $360^\circ$ videos that can be $4 \times$ to $6 \times$ larger than regular videos under the same perceived quality \cite{qian2018flare}. Moreover, streaming the sheer volume of $360^\circ$ videos to remote data centers or clouds over the dynamic public Internet \cite{zhang2018awstream} may incur excessive bandwidth costs and unacceptable delays \cite{guan2019pano}. As a result, an edge-assisted architecture, in which all or part of the analytics workloads are offloaded to edge servers (or edge clouds) for processing, is necessary to deliver an ideal solution.


There remain, however, distinct challenges toward edge-assisted immersive analytics for $360^\circ$ videos. A $360^\circ$ frame is internally represented as a spherical image, which is projected onto a $2$D plane to enable storage and delivery in practice \cite{hu2017deep, cubemap-facebook, sun2017learning}. Yet, the projected panoramic image cannot be simply treated as a regular video frame. This is because typical off-the-shelf vision models are designed for and trained on $2$D perspective images (PIs). The sphere-to-plane projections inevitably introduce geometric distortions and border discontinuities and thus hurt the accuracy of these models \cite{sun2017learning, coors2018spherenet, chou2020360}. This precludes most vision models from being used for high-accuracy panoramic image analysis. Furthermore, even with vision models specifically designed for panoramic images \cite{coors2018spherenet, lee2019spherephd}, the bandwidth costs of uploading high-resolution (e.g., $8$K) panoramic frames to edge servers can hardly be afforded by today's network infrastructures \cite{qian2018flare, guan2019pano}. One workaround is to project the entire spherical content to multiple distortion-free PIs \cite{yang2018object, eder2020tangent}. However, reducing the latency of analyzing all PIs without compromising accuracy remains a problem.

In this paper, we present \texttt{OmniSense}, an \emph{edge-assisted} immersive video analytics framework that achieves low latency and high accuracy by adaptively utilizing vision models with distinct resource demands and capabilities to analyze different spherical regions of interest (SRoIs). Essentially, \texttt{OmniSense} is empowered by off-the-shelf vision models and does not require redesigns or retraining of existing models. Thus, advances in vision models designed for regular videos can also benefit $360^\circ$ videos. To the best of our knowledge, \texttt{OmniSense} is the first immersive video analytics framework with a special focus on resource efficiency in practical systems. Our contributions can be summarized as follows.
\begin{itemize}
    \item By analyzing a $360^\circ$ video dataset collected from the real world, we identify the content characteristics and the resource-saving opportunities of $360^\circ$ videos.
    \item We propose \texttt{OmniSense}, an edge-assisted framework that maximizes the overall accuracy under computational power, network, and latency constraints by dynamically and adaptively allocating different vision models to analyze the PIs corresponding to different SRoIs.
    \item We design a lightweight SRoI prediction algorithm and a content-specific model performance estimation method. Based on them, we further solve a latency-constrained model allocation problem.
    \item We implement and deploy a prototype of \texttt{OmniSense} with commodity devices for performance evaluation. Extensive evaluation results demonstrate that it improves the accuracy of baselines by up to $114.6\%$ with similar end-to-end latencies and approximately hits or exceeds the highest accuracy of baselines with $2.0\times$ -- $2.4\times$ speedups.
\end{itemize}

\section{Background and Related Work}
\subsection{Online Analytics for Regular Videos}

Online video analytics \cite{ananthanarayanan2017real} has attracted a lot of attention in recent years due to the increasing deployments of networked cameras and the advances in computer vision algorithms. A variety of systems \cite{zhang2017live, jiang2018chameleon, wang2019bridging, guo2021crossroi} have been designed to automatically analyze the content of regular videos with vision models. Early systems depend on resource-rich data centers or clouds to realize high-accuracy analytics but need the support of dedicated or high-quality network links \cite{zhang2017live, jiang2018chameleon}. To facilitate video analytics for cameras with wireless or unstable networks, edge resources have been examined by existing systems \cite{jiang2018mainstream, guo2021crossroi, jiang2021flexible}. Despite the reduced network delays, resource-constrained edge devices hardly achieve high accuracy without the help of powerful backends \cite{ananthanarayanan2017real, zhang2021elf}.

Thus, collaborating geo-distributed resources (e.g., device, edge, and cloud resources) is considered promising to achieve low latency and high accuracy \cite{ananthanarayanan2017real}. Several techniques have emerged in this context to improve accuracy while reducing the amount of data transferred between front-end devices and backends. One popular technique is to adjust video encoding knobs (e.g., resolution) together with the task placement knob (e.g., front-end devices or backends) to strike a balance between resource and accuracy \cite{ran2018deepdecision}. Given that events of interest can be sparse in time, video frame filtering \cite{canel2019scaling} is explored to save resources by filtering out irrelevant video frames on front-end devices and only uploading event-related frames to backends for further processing. Additionally, to overcome the resource challenges of accommodating large DNN inference on weak front-end devices, splitting the model inference across front-end devices and backends to minimize the intermediate data transfer overhead is also studied \cite{emmons2019cracking}. Recently, pruning redundant information in video frames based on regions of interest (RoIs) has proven to be another viable way to reduce the amount of data being transferred and analyzed \cite{du2020server, zhang2021elf}.

\begin{figure}[!t]
    \centering
    \includegraphics[width=\linewidth]{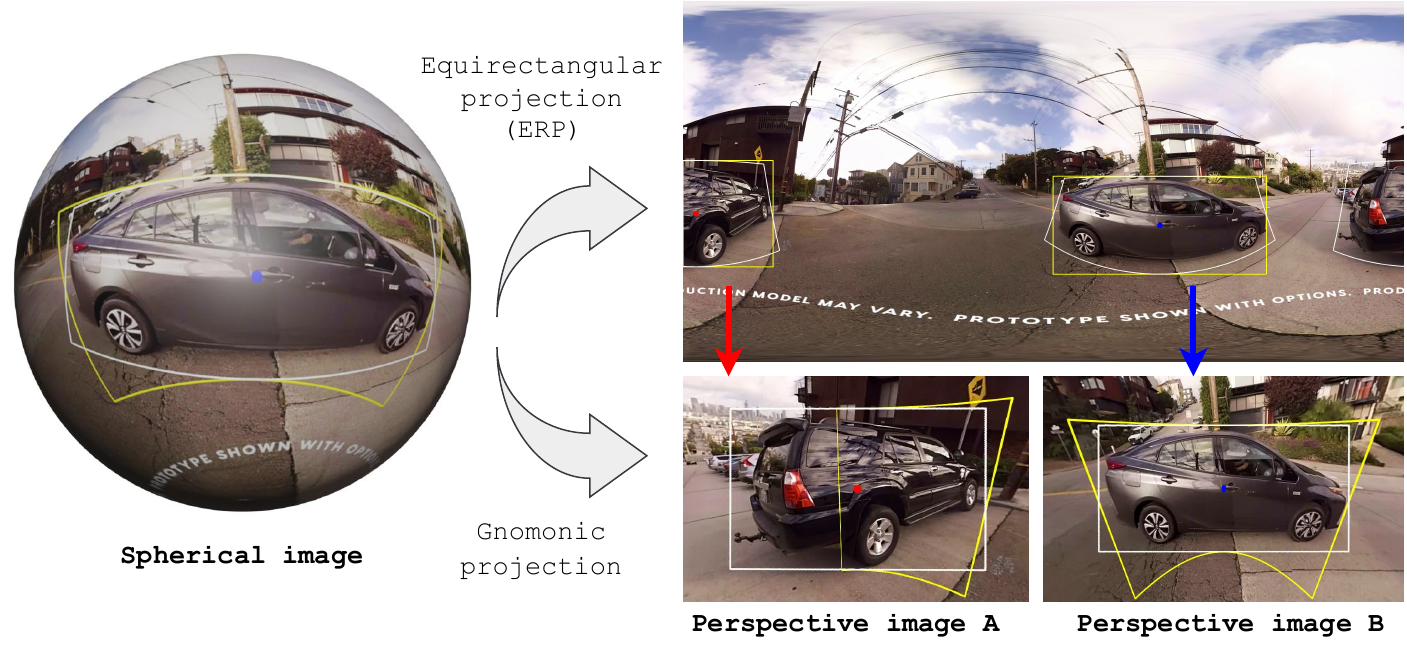}
    \caption{An illustrative example of spherical image projections and spherical criteria. The yellow boxes are rectangular BBs, and the white boxes are SphBBs (image source: \cite{yang2018object}).}
    \label{fig:360-projection}
    \vspace{-0.3cm}
\end{figure}

\subsection{Immersive Video Analytics}

Several methods have been proposed to project a spherical frame captured by omnidirectional cameras onto a $2$D plane, such as equirectangular projection (ERP) \cite{hu2017deep} and Cubemap projection \cite{cubemap-facebook}. Accordingly, one natural idea to implement immersive video analytics is directly applying off-the-shelf vision models to the projected panoramic images \cite{hu2017deep}. \emph{Yet, the projections unavoidably introduce geometric distortions and discontinuities in panoramic images, negatively impacting the accuracy of vision models initially designed for and trained on flat images \cite{sun2017learning, coors2018spherenet, chou2020360}.} One example is illustrated in Fig.~\ref{fig:360-projection}. The upper-right $2$D image, which is obtained from the left spherical image via ERP, suffers from severe distortions in the polar regions and discontinuities on the boundary.

One workaround is to denote the entire spherical content as multiple distortion-free PIs via general perspective projections, e.g., gnomonic projection \cite{coors2018spherenet} shown in Fig.~\ref{fig:360-projection}. Each projected PI corresponds to a partial FoV and can be well analyzed by off-the-shelf vision models. For instance, Eder \emph{et al.} \cite{eder2020tangent} propose to represent a spherical image as multiple planar image grids tangent to a subdivided icosahedron and then apply existing DNNs on these tangent images. \emph{Since it requires processing a large number of PIs to minimize distortions and cover the entire spherical content, this strategy is resource-intensive and time-consuming.} Yang \emph{et al.} \cite{yang2018object} present a multi-projection method that partitions the sphere into four wide overlapping sub-windows and maps each sub-window to a plane by stereographic projection. \emph{They focus on improving accuracy and do not address the computation and network resource challenges in implementing practical systems.}

Considering that convolutional neural networks (CNNs) are widely adopted as feature extractors in popular vision models, another line of research focuses on explicitly encoding invariance against the projection distortions into CNNs \cite{sun2017learning, coors2018spherenet}. For example, Sun \emph{et al.} \cite{sun2017learning} propose to increase convolutional kernel size towards the polar regions to approximate the distortions in ERP images. Coors \emph{et al.} \cite{coors2018spherenet} instead propose to process ERP images by adapting the sampling grid locations of convolutional filters based on the geometry of the spherical image representation. All these studies require redesigning the convolutional filters and retraining (or at least fine-tuning) the existing models. \emph{In contrast, we work toward a plug-and-play underlying analytic framework for immersive videos, i.e., applying existing models trained on PIs without modifications.}

\section{Motivation Study}
In this work, we take one fundamental video analytics task, \emph{object detection} \cite{wang2021scaled}, as a case study to investigate the content characteristics of $360^\circ$ videos and motivate the design of resource-efficient immersive video analytics systems.

\begin{table}[!t]
\centering
\begin{tabularx}{0.47\textwidth}{rrrr}
\toprule
Immersive Video Name & Video Source & Resolution & Frames \\
\midrule \midrule
New-Orleans-drive & YouTube \cite{new-orleans-driving} & $7680 \times 3840$ & $2100$ \\
Expressway-drive & YouTube \cite{expressway-driving} & $5760 \times 2880$ & $2100$ \\
Chicago-drive & YouTube \cite{chicago-driving} & $7680 \times 3840$ & $2100$ \\
Sunny-walk1 & self-captured & $5376 \times 2688$ & $2100$ \\
Sunny-walk2 & self-captured & $5376 \times 2688$ & $2100$ \\
Cloudy-walk & self-captured  & $5376 \times 2688$ & $2100$ \\
\bottomrule
\end{tabularx}
\caption{Summary of our $360^\circ$ video dataset.}
\label{tab:video-dataset}
\end{table}

\subsection{Motivation Study Setup}
\label{sec:motivation-setup}

\noindent \textbf{Video dataset and models:} We use a self-collected ultra-high-definition $360^\circ$ video dataset shown in TABLE~\ref{tab:video-dataset} to cover various real-world scenarios. The YouTube videos are captured by omnidirectional cameras mounted on the roof of cars driving through cities or expressways in the US. The self-shot videos are captured by a GoPro MAX camera handheld by a person walking through blocks or campus under different illumination conditions. Each video has a length of $7$ minutes and is stored in the ERP format. In this work, we employ the scaled-YOLOv$4$ \cite{wang2021scaled} for the object detection task since it provides a set of model variants with varying resource demands and accuracies, and the details are shown in TABLE~\ref{tab:models}. The model weights are pre-trained on the MS COCO dataset\cite{lin2014microsoft}. Generally, a model variant with a higher input size achieves a higher accuracy at the cost of higher resource consumption. Therefore, one can strike a good balance between resource and accuracy by choosing appropriate model variants.

\begin{table}[!t]
\centering
\begin{tabularx}{0.48\textwidth}{rrrr}
\toprule
Model Name (Index) & Model Size & Input Size & Location\\
\midrule \midrule
YOLOv$4$-Tiny-$416$ ($1$) & $23$ MB & $416 \times 416$ & Mobile Device \\
YOLOv$4$-CSP-$512$ ($2$) & $202$ MB & $512 \times 512$ & Edge Server \\
YOLOv$4$-CSP-$640$ ($3$) & $202$ MB & $640 \times 640$ & Edge Server \\
YOLOv$4$-P$5$ ($4$) & $271$ MB & $896 \times 896$ & Edge Server \\
YOLOv$4$-P$6$ ($5$) & $487$ MB & $1280 \times 1280$ & Edge Server\\
\bottomrule
\end{tabularx}
\caption{Summary of model variants used in this work.}
\label{tab:models}
\end{table}

\begin{figure*}[!t]
  \begin{minipage}{0.325\linewidth}
        \centering
        \includegraphics[width=\linewidth]{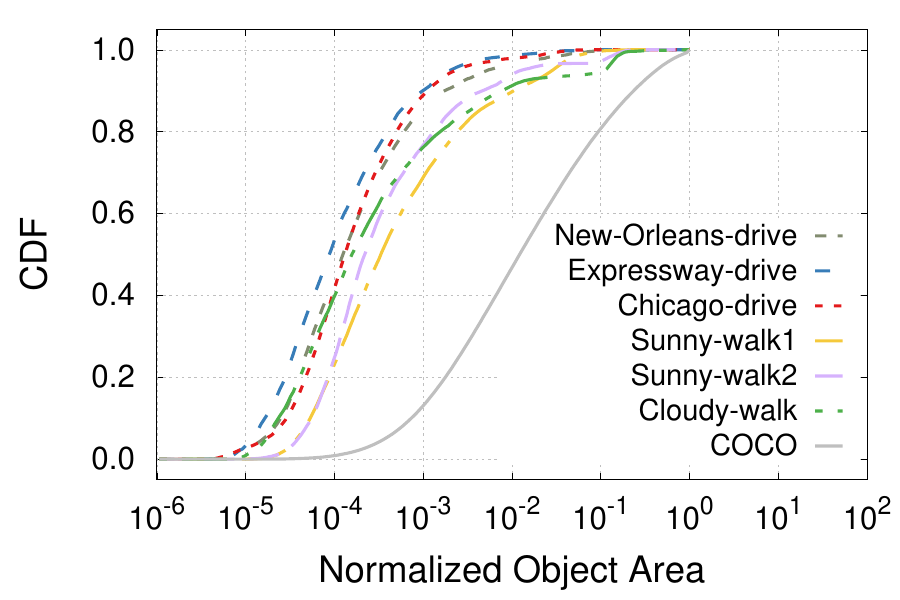}
        \caption{CDFs of NOA for $360^\circ$ videos in our dataset and $2$D images in COCO.}
        \label{fig:object-area-cdfs} 
  \end{minipage}\hfill
  \begin{minipage}{0.325\linewidth}
        \centering
        \includegraphics[width=\linewidth]{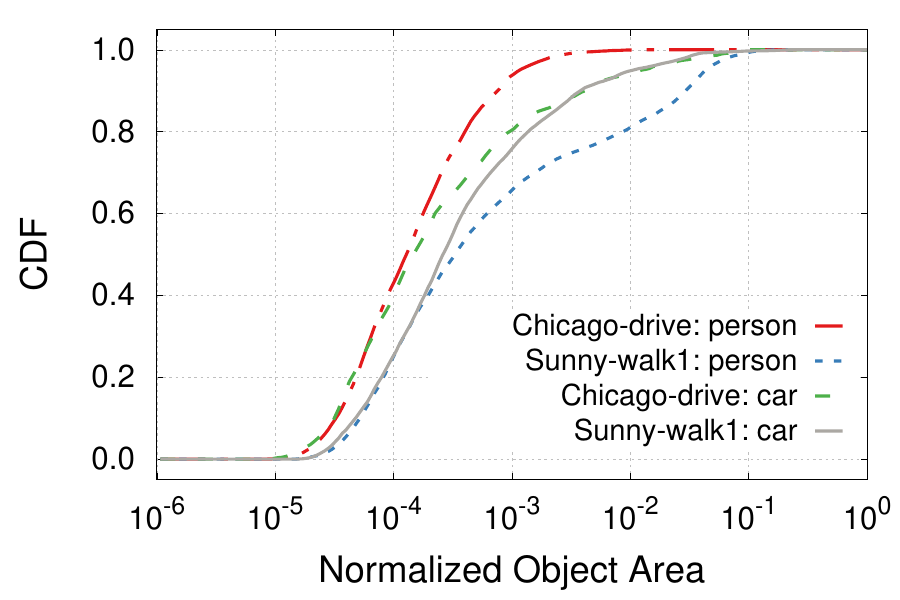}
        \caption{CDFs of NOA for two specific object categories.}
        \label{fig:category-area-cdfs}
  \end{minipage}\hfill
  \begin{minipage}{0.325\linewidth}
        \centering
        \includegraphics[width=\linewidth]{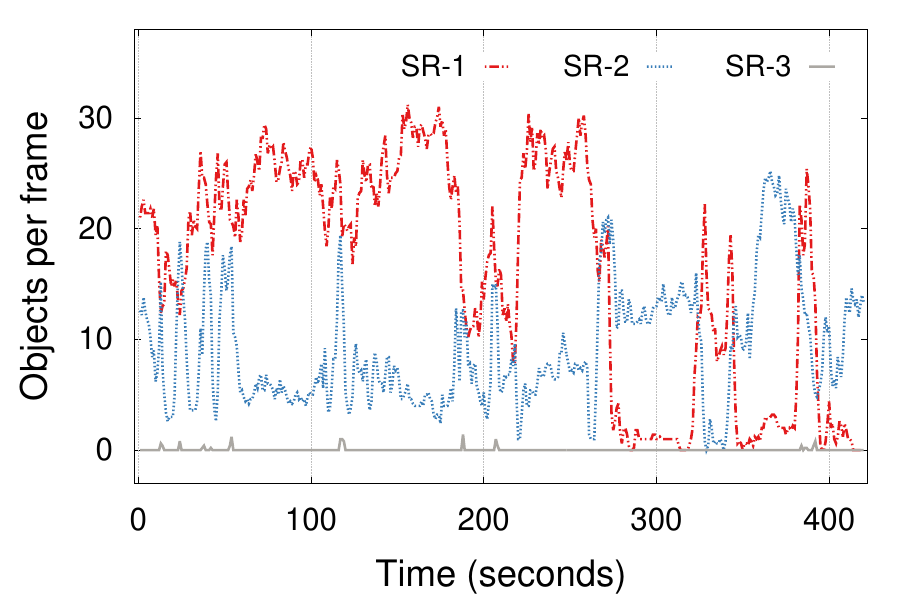}
        \caption{Object number variations in different SRs (video: New-Orleans-drive).}
        \label{fig:fovs-aopf-vary}
  \end{minipage}\hfill
\end{figure*}

\noindent \textbf{Immersive object detection criteria:} Object detection aims to find out the location and category of objects in an image. For $2$D images, the locations of detected objects are typically annotated by rectangular bounding boxes (BBs). Unfortunately, as illustrated by Fig.~\ref{fig:360-projection}, due to the particular geometry of spherical images, rectangular BBs (in yellow) on the ERP image fail to bound objects on a sphere tightly and precisely. Such BBs experience distortions on PIs tangent to the object centers as well. As a result, several $360^\circ$ object detection criteria have been proposed to bound objects and calculate the intersection-over-union (IoU) of two objects, e.g., BBs on tangent planes \cite{coors2018spherenet}, and circle BBs and IoUs \cite{lee2019spherephd}.

In this work, we use the spherical criteria, including spherical BB (SphBB) and spherical IoU (SphIoU), proposed in \cite{zhao2020spherical} as it is fast and accurate. The white boxes in Fig.~\ref{fig:360-projection} show what the SphBBs look like on the spherical image, ERP image, and PIs. Specifically, a SphBB is directly defined on a sphere and represented by a spherical region $(\theta, \phi, \Delta_{\theta}, \Delta_{\phi})$, where $\theta$ and $\phi$ denote the longitude and latitude of the object center, respectively; $\Delta_{\theta}$ and $\Delta_{\phi}$ denote the horizontal and vertical FoVs of the object's occupancy, respectively. All values are in degrees/radians.

\noindent \textbf{Ground-truth immersive object detection results:} Lacking of fast and scalable $360^\circ$ video annotation tools, we have developed our own annotation pipeline to generate approximate ground-truth results offline. Since maximizing accuracy is the first concern in ground-truth annotation, we have employed the most accurate model, namely YOLOv4-P6, for this purpose.

We first run the model with a low confidence threshold (e.g., $0.3$) to obtain rough detection results from the input ERP frame. Given that an object's center is nearly consistent across various spherical image representations \cite{yang2018object}, we extract distortion-free PIs centered at the detected object centers via gnomonic projection. Concretely, for each detected object, we project a $60^\circ \times 60^\circ$ spherical region centered at the object center to a plane, where the vision model is applied for further precise refinement. We only keep the detection results of objects entirely enclosed by a PI to avoid repetitive detections at the boundary. Then, the fine-grained detection results of the PIs are back-projected to the sphere, and we obtain the SphBB for each detection by spherical coordinate transformations. Finally, spherical non-maximum suppression (NMS) \cite{zhao2020spherical} is applied to all obtained SphBBs to produce the final results.

We empirically set the projected spherical region to $60^\circ \times 60^\circ$ to eliminate distortions while covering as many objects as possible. For rare objects close to the camera, spanning a wide FoV greater than $60^\circ$, we manually annotate them. The intuition behind this approach is that due to the increased input resolution, small objects that cannot be detected from the panoramic image are likely to be detected from PIs. We examined the detection results and found that this method is approximately accurate, although it may occasionally ignore tiny standalone objects.

\subsection{Observations and Implications}

\noindent \textbf{Most objects in $360^\circ$ videos occupy only a tiny area of a frame.}
We first examine the visible size of objects in $360^\circ$ videos. To quantify the object size, we analyze the ground-truth detection results of all videos in our dataset. Particularly, we calculate the area of each SphBB on a unit sphere\footnote{The area of a SphBB can be calculated by $2 \Delta_{\theta} \sin (\Delta_{\phi} / 2)$ \cite{zhao2020spherical}.} and normalize the SphBB area by the surface area of the sphere. Fig.~\ref{fig:object-area-cdfs} shows the cumulative distribution functions (CDFs) of normalized object area (NOA) for all $360^\circ$ videos. For comparison, we also plot the CDF of the NOA for COCO's training set\footnote{The normalized object area (NOA) for $2$D images is calculated by dividing the area of the object's rectangular BB by the area of the whole image.}. As shown, the area occupied by objects in $360^\circ$ frames is much smaller than that in conventional $2$D images, suggesting that processing a $360^\circ$ frame as a whole requires vision models capable of detecting tiny objects in high-resolution input frames. Unfortunately, such models are known to be slow and expensive \cite{jiang2021flexible}. The status-quo strategy, i.e., downsampling the input frame to match the input size of DNNs, makes tiny objects ambiguous, hence leading to poor detection accuracy. For instance, if we downsample a $360^\circ$ frame into $640 \times 640$ pixels, a considerable portion of objects with a normalized area between $10^{-5}$ to $10^{-4}$ will only take an area from $4$ to $41$ pixels, which is too small to be detected.

\emph{The observations reveal the gap between the requirements of spherical frames and the capability of the off-the-shelf vision models. One prospective way to bridge this gap is employing multiple models to analyze the entire spherical content in a ``divide-and-conquer'' manner.}

\noindent \textbf{The area of objects in the same category differs by several orders of magnitude.} Fig.~\ref{fig:category-area-cdfs} further displays the CDFs of NOA for two particular categories: person and car. As can be seen, the area occupied by objects of the same category can differ by three to four orders of magnitude. For the \texttt{Chicago-drive} video, cars show intenser variations in size than people, as people usually appear on the sidewalks at a certain distance from the camera mounted on the roof of the driving car. This is different from the \texttt{Sunny-walk1} video, where people can be very close to the handheld camera, thus occupying a large portion of the captured view.

\emph{The observations suggest that the object size distribution is video-specific, and one cannot simply tell the object category from its visible size. Both object size and object category are crucial reference factors in characterizing video content and the capability of detection models.}

\noindent \textbf{The spatial distribution of objects in $360^\circ$ frames is biased.} We next investigate the spatial distributions of objects in $360^\circ$ frames. Specifically, we consider three spherical regions (SRs), each covering a $60^\circ \times 60^\circ$ FoV on the sphere. They are denoted as SR-$1$ $(0, 0, 60, 60)$, SR-$2$ $(-90, 0, 60, 60)$, and SR-$3$ $(0, 90, 60, 60)$, respectively. We count the number of objects whose centers fall into these SRs and show the variations in Fig.~\ref{fig:fovs-aopf-vary}. As shown, the spatial distribution of objects is biased, and there are massive pixels without useful information for the object detection task. For example, the SR-$3$, which captures a view of the sky, contains rare objects of interest. As a result, treating SR-$3$ the same as SR-$1$ and SR-$2$ would result in a waste of resources. Furthermore, as verified in existing studies \cite{jiang2018chameleon, jiang2021flexible}, leveraging high-accuracy models to process content with simple scenes only leads to marginal accuracy gain.

\emph{The observations mean that despite the large size of $360^\circ$ frames, many pixels can be pruned to save resources, and using models with different capabilities to handle different spherical regions can be a promising way to further improve resource efficiency.}

\noindent \textbf{The content of $360^\circ$ videos can be highly dynamic.} Another observation from Fig.~\ref{fig:fovs-aopf-vary} is that for both SR-$1$ and SR-$2$, the number of objects varies substantially over time. This is primarily due to changes in the scene as the camera moves. For example, objects can frequently appear or disappear as the car that the camera mounted turns or drives through intersections. This indicates that even for the same SR, the most suitable vision model may change over time. 

\emph{This observation implies that an ideal resource allocation scheme should be able to adapt to the variations in $360^\circ$ video content to maximize resource efficiency.}

\begin{figure}[!t]
    \centering
    \includegraphics[width=\linewidth]{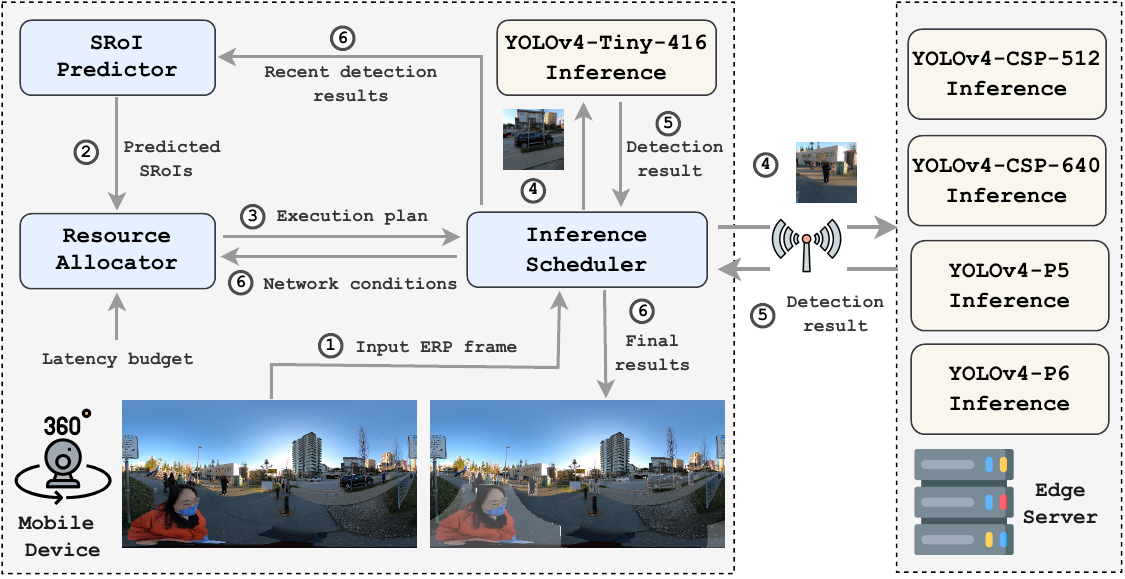}
    \caption{Overview of \texttt{OmniSense}.}
    \label{fig:system-overview}
\end{figure}

\section{System Design}

Our motivation study has revealed the resource-saving opportunities in immersive video analytics. Nevertheless, several challenges must be overcome to implement a resource-efficient system. To prune useless pixels in $360^\circ$ frames and facilitate the application of off-the-shelf vision models, we introduce the \emph{spherical region of interest (SRoI)} to describe an SR containing objects of interest. Immediately, one challenge that arises is how to identify SRoIs. This is not an easy one-time job since the set of SRoIs is vulnerable to video scene changes. Moreover, selecting the most suitable inference model for an SRoI is tricky as it requires insights into the SRoI's content characteristics, the model's capability, and even the network conditions if the remote inference is enabled. In addition, high-level applications tend to impose a fixed analysis latency budget on one frame, regardless of the number of its SRoIs. This makes the model choices of distinct SRoIs coupled since allocating unnecessarily excessive resources to one SRoI may deprive other SRoIs of the opportunity to improve accuracy.

We address the challenges by presenting \texttt{OmniSense}, an \emph{edge-assisted} framework for low-latency and high-accuracy immersive video analytics. \texttt{OmniSense} hypothesizes that the analytics workloads can be offloaded from a mobile device capturing $360^\circ$ videos to an edge server, which is physically close to the mobile device. The mobile device runs cheap models to analyze simple SRoIs and provide a minimum accuracy guarantee when the network condition is poor, while the edge server executes expensive models to analyze complex SRoIs with improved accuracy in good network conditions.

Fig.~\ref{fig:system-overview} shows an overview of \texttt{OmniSense}. Specifically, for each input ERP\footnote{Although we use ERP as the input frame format in this work, \texttt{OmniSense} in principle could be applied to other $360^\circ$ image representations as the bounding boxes and computations are defined directly on the sphere.} frame \circled{1}, the \textbf{SRoI predictor} (\cref{sec:roi-prediction}) first predicts where the objects of interest are likely to appear based on the detection results of the most recent frames. It then feeds the coordinates and content characteristics vectors of the predicted SRoIs to the \textbf{resource allocator} \circled{2}, which estimates the detection accuracy and inference latency of each model on each SRoI (\cref{sec:model-estimate}). Given the analysis latency budget and current network conditions, the \textbf{resource allocator} further solves a model allocation problem and outputs an execution plan for the input frame (\cref{sec:model-allocation}). Once the execution plan is received \circled{3}, the \textbf{inference scheduler} extracts the PI for each SRoI from the input ERP frame via gnomonic projection, and the size of each PI is the input size of its allocated model. After projection, PIs are immediately sent to appropriate local or remote servers for inference \circled{4}. Subsequently, the inference results of all PIs \circled{5} are integrated and transformed to obtain the final spherical detection results, which will be further sent back to the \textbf{SRoI predictor} for the SRoI prediction of the next frame, and the latest network conditions will also be updated to the \textbf{resource allocator} \circled{6}.

\subsection{Lightweight SRoI Prediction}
\label{sec:roi-prediction}

\begin{algorithm}[!t]
\SetAlgoLined
\small
\KwData{$f$; $\gamma$; $\mathcal{O}$ (detected objects of the most recent $\delta$ frames)}
\KwResult{A set of predicted SRoIs $\mathcal{R}$}
Initialize SRoI sets $\mathcal{S} \gets \mathcal{\emptyset}$, $\mathcal{S'} \gets \mathcal{\emptyset}$ \;
Get the number of all historical objects $N \gets |\mathcal{O}|$ \;
\ForEach{object $o \in \mathcal{O}$}{
    \If{$o$ can be covered by an $f \times f$ FoV}
    {
        $merged \gets False$ \;
        \ForEach{SRoI $s \in \mathcal{S}$}{
            $hFoV, vFoV \gets$ merged horizontal and vertical FoVs for the set $s.objects \cup \{o\}$\;
            \If{$hFoV < f$ \textbf{and} $vFoV < f$}{
                $s.objects \gets s.objects \cup \{o\}$ \;
                $s.FoV \gets (hFoV, vFoV)$ \;
                $merged \gets True$; $break$ \;
            }
        }
        \If{$not~merged$}{
            $new\_s \gets$ create a new SRoI with $o$ \;
            $\mathcal{S} \gets \mathcal{S} \cup \{new\_s\}$ \;
        }
    }
    \Else{
        Create a new special SRoI $s'$ with $o$ \;
        $s'.center \gets o.center$; $s'.FoV \gets$ $\gamma \times o.FoV$ \;
        Calculate content characteristics $s'.ccv$ based on $o$ \;
        $s'.\alpha \gets 1~/~N$ \;
        $\mathcal{S'} \gets \mathcal{S'} \cup \{s'\}$ \;
    }
}
\ForEach{SRoI $s \in \mathcal{S}$}{
    Calculate SRoI center $s.center$ according to $s.FoV$ \;
    Calculate $s.ccv$ based on $s.objects$\;
    $s.\alpha \gets |s.objects|~/~N$ \;
    $s.FoV \gets (f, f)$ \;
}
$\mathcal{R} \gets \mathcal{S'} \cup \mathcal{S}$\;
\Return $\mathcal{R}$ \;
\caption{SRoI Prediction Algorithm}
\label{alg:SRoI-prediction}
\end{algorithm}

In spite of the intense content variations in a $360^\circ$ video, consecutive frames have the smallest content differences. This motivates us to use the detection results of the most recent frames to acquire up-to-date knowledge about the spatial distribution of objects. Similar to the SphBB, we use the tuple $(\theta, \phi, \Delta_{\theta}, \Delta_{\phi})$ to denote an SRoI. Although one SRoI can be of any size, we assume that the horizontal and vertical FoVs of all SRoI are $f$ to simplify the subsequent resource allocation. Then, based on the historical detection results of the most recent $\delta$ ($2$ by default) frames, we design a lightweight prediction algorithm, as shown in Algorithm \ref{alg:SRoI-prediction}, to obtain a set of SRoIs covering all historical objects.

The key idea is that for each historical object, we try to merge it into an existing SRoI; if it fails, i.e., the merged horizontal or vertical FoV exceeds $f$, we will create a new SRoI to enclose it. Setting an appropriate value for $f$ is tricky. A small value may cut large objects apart, while a large value may introduce projection distortions, both of which will lead to inaccurate detection results. As such, we empirically set $f$ to $60^\circ$ to avoid projection distortions and then deal with large objects in a particular way. Specifically, we create a special SRoI with an area $\gamma \times$ ($1.1$ by default) the area of one object if it cannot be enclosed by an $f \times f$ FoV. Since the special SRoI can be very large, which may introduce projection distortions to co-located small objects, we only keep the detection result of the largest object for it.

Relying solely on the historical detection results can have adverse cascading effects. For example, with a tight latency budget, only simple models could be used for latency reduction. This may cause only a few objects being detected, thereby reducing the number of SRoIs predicted for subsequent frames. The reduced SRoIs may further reduce the number of detected objects for subsequent frames. To break the vicious circle, we design a \emph{spherical object discovery} mechanism. It opportunistically exploits the underutilized latency budget to explore new objects by sending an ERP frame to the server for inference. The detection results will be converted to SphBBs and appended to the historical detection results for the SRoI prediction of the next frame. This mechanism will be triggered automatically if the number of predicted SRoIs is consistently low. The rationale behind this is that although ERP detection cannot precisely discover and locate all objects, it can help discover new spherical regions where objects appear globally.

\subsection{Content-Specific Model Performance Estimation}
\label{sec:model-estimate}

To allocate suitable vision models for the predicted SRoIs, we need to estimate each model's performance for each SRoI. For one-shot object detection models like scaled-YOLOv$4$, the model inference latency on a resource-fixed device is approximately consistent and can be profiled offline. We thus estimate the per-image inference latency of an object detection model as its average inference time across thousands of runs on the target device.

Due to the bias in the training dataset, the accuracy of a pre-trained object detection model varies with image content. A recent work \cite{jiang2021flexible} considered stationary cameras and attempted to estimate a model's accuracy for a $2$D image by integrating its detection capabilities at different object size levels with the object size distribution of the image. However, as our measurement results suggest, object size alone is insufficient to characterize $360^\circ$ video content captured by moving cameras. Therefore, we consider both object size and category to design an accuracy estimation method tailored to our problem.

We first group objects into three size levels: small, medium, and large. Instead of using pixels as in \cite{jiang2021flexible}, we utilize the NOA as a unified scale measure to correlate object size in $2$D and spherical images. Since our models are trained on COCO, we set the size level thresholds to the $33.33$ percentile ($0.0044$) and the $66.66$ percentile ($0.0354$) of COCO's NOA distribution, i.e., small objects have NOAs not greater than $0.0044$, and medium objects have NOAs between $0.0044$ and $0.0354$. Based on this classification, we formally define the \emph{general accuracy vector (gav)} for model $i$ as
\begin{equation}
    \mathbf{A_i} = [a_i^{s1}, \cdots, a_i^{sn}, a_i^{m1}, \cdots, a_i^{mn}, a_i^{l1}, \cdots, a_i^{ln}]
\end{equation}
where $a_i^{sc}$, $a_i^{mc}$, and $a_i^{lc}$ denote model $i$'s accuracy in detecting small, medium, and large objects of a particular category $c$, respectively; $n$ is the number of object categories. $\mathbf{A_i}$ can be estimated by offline profiling publicly available datasets. This work uses COCO, whose category number $n$ is $80$.

Meanwhile, the dynamic content of an SRoI can be characterized by its object size and category distribution as well. We formally define the \emph{content characteristics vector (ccv)} for one SRoI $j$ as
\begin{equation}
    \mathbf{P_j} = [p_j^{s1}, \cdots, p_j^{sn}, p_j^{m1}, \cdots, p_j^{mn}, p_j^{l1}, \cdots, p_j^{ln}]
\end{equation}
where $p_j^{sc}$, $p_j^{mc}$, and $p_j^{lc}$ represent the occurrence probabilities of objects belonging to a specific category $c$ at small, medium, and large size levels, respectively. As shown in Algorithm~\ref{alg:SRoI-prediction} (line $18$ and line $23$), the \textbf{SRoI predictor} estimates $\mathbf{P_j}$ for each predicted SRoI $j$ based on its absorbed historical objects. For instance, $p_j^{sc}$ is estimated as the frequency of small objects of category $c$ in all historical objects merged by SRoI $j$. With the \emph{gav} and \emph{ccv}, the detection accuracy of model $i$ on SRoI $j$ can be estimated by $\mathbf{A_i} \cdot \mathbf{P_j}$. In this way, we can estimate the inference latency and detection accuracy of each candidate model for each predicted SRoI, which will guide us to allocate more resources to SRoIs that bring more accuracy benefits.

\subsection{Latency-Constrained Model Allocation}
\label{sec:model-allocation}

Given the \emph{analysis latency budget} $T$ for analyzing a $360^\circ$ frame, the \textbf{resource allocator} is responsible for distributing the limited resources across the predicted SRoIs to achieve a high overall detection accuracy. Assume that we have a set of models $\mathcal{M} = \{0,1,\cdots,m\}$ and a set of predicted SRoIs $\mathcal{R} = \{1, 2, \cdots, r\}$. Let $\mathit{x_{i,\,j}}$ be a $0$-$1$ indicator variable that gets $1$ if model $i$ is allocated to analyze SRoI $j$. We set $\mathit{x_{0,\,j}}$ to $1$ to represent a special case where SRoI $j$ will be ignored without processing. This happens when the SRoI contains very little visual information or when the available resources are exceptionally limited.

The goal of resource allocation is to find the optimal execution plan, i.e., the optimal set $\mathcal{X} = \{\mathit{x_{i,\,j}}\ |~i \in \mathcal{M}, j \in \mathcal{R} \}$, so that the overall detection accuracy can be maximized under the analysis latency budget. Since SRoIs with varying numbers of objects can contribute differently to the overall detection accuracy, we introduce $\mathit{A_{i,\,j}} = \alpha_j \cdot \mathbf{A_i} \cdot \mathbf{P_j}$ to denote the weighted accuracy achieved by allocating model $i$ to SRoI $j$, where $\mathit{\alpha_j}$ is the probability of an object appearing in SRoI $j$. It can be estimated as the ratio of the number of historical objects merged by SRoI $j$ to the number of all historical objects (line $19$ and line $24$ in Algorithm~\ref{alg:SRoI-prediction}). Then, the latency-constrained model allocation problem can be formally expressed as

\begin{equation}
\begin{aligned}
    &\max\limits_{\mathit{x_{i,\,j}}} \quad
    \sum_{j \in \mathcal{R}}\sum_{i \in \mathcal{M}} \mathit{A_{i,\,j}} \cdot \mathit{x_{i,\,j}}\\
    & s.t. \quad
    \begin{cases}
    \mathcal{L} (\mathcal{X}) \leq T, \quad \mathcal{X} = \{\mathit{x_{i,\,j}}\ |~i \in \mathcal{M}, j \in \mathcal{R} \}\\
    \sum_{i \in \mathcal{M}} \mathit{x_{i,\,j}} = 1, \quad \forall j \in \mathcal{R} \\
    \mathit{x_{i,\,j}} \in \{0,1\}, \quad \forall i \in \mathcal{M}, \quad \forall j \in \mathcal{R} \\
    \end{cases}
\end{aligned}
\label{equ:opt}
\end{equation}
where $\mathcal{L}(\mathcal{X})$ is the \emph{analysis latency} of the execution plan $\mathcal{X}$. We next introduce how to estimate $\mathcal{L}(\mathcal{X})$ in detail.

\begin{figure}[!t]
    \centering
    \includegraphics[width=0.9\linewidth]{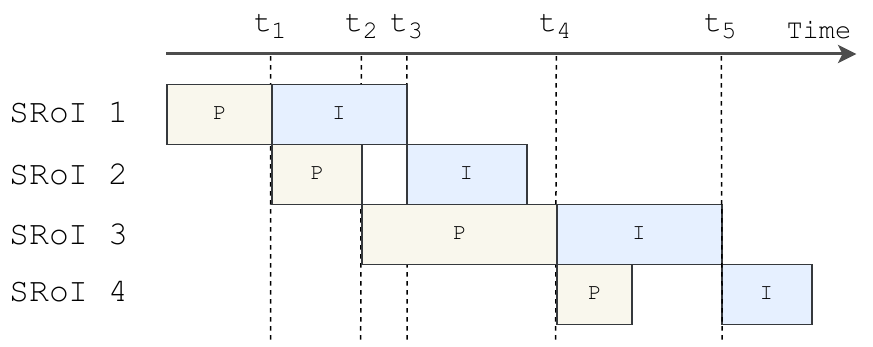}
    \caption{An illustrative example of pipelining preprocessing (P) and inference (I).}
    \label{fig:pipeline}
    \vspace{-0.5cm}
\end{figure}

Let $\mathit{d_{i,\,j}}$ denote the latency of analyzing SRoI $j$ with model $i$, and it can be divided into two parts: the \emph{preprocessing delay} and the \emph{inference delay}. The \emph{preprocessing delay} consists of the projection time for the mobile device to acquire the PI of the SRoI from the input frame and the optional encoding time taken to compress the PI for network delivery (only for remote inference). The projection and compression time of an SRoI is related to the resolution of its projected PI. For example, projecting an SRoI to a PI of $1280 \times 1280$ takes more time than projecting it to a PI of $512 \times 512$. The resolution of each SRoI's PI is set as the input size of its allocated model to avoid resizing the image. Furthermore, PIs are compressed in a lossless format by default to prevent accuracy loss caused by the degraded image quality. Since there are $m$ models, the resolution of PIs has $m$ options. For each option, we profile the projection and compression delays of SRoIs on the target device offline. Then, we can estimate the \emph{preprocessing delay} of each candidate model for each SRoI.

The \emph{inference delay} includes the network delivery delay and the model inference time. The former for local model inference is $0$. For remote model inference, we use an online passive profiling method similar to that in \cite{zhang2021elf}. Specifically, the server calculates the mean network delivery delays of the most recent $\omega$ ($7$ by default) requests for each model and synchronizes with the mobile device by piggybacking the updated network conditions on the detection results. The model inference time can be estimated with the method mentioned in \cref{sec:model-estimate}.

If the predicted SRoIs are processed one by one in the order of preprocessing and inference, the problem (\ref{equ:opt}) can be considered as a variant of the \emph{multiple-choice knapsack problem} \cite{pisinger1995minimal}. Unfortunately, such a serial scheme may introduce significant latency, especially for weak mobile devices. We thus propose a \emph{pipelining preprocessing and inference} technique to accelerate the end-to-end processing. The key enabler is that there are no computational dependencies between SRoIs. The inference of one SRoI and the preprocessing of the next SRoI can be naturally pipelined. Fig.~\ref{fig:pipeline} shows an illustrative example, where the preprocessing of one SRoI starts immediately after the preprocessing of the previous SRoI completes, i.e., at $t_1$, $t_2$, and $t_4$. The SRoI inference starts when the preprocessing is completed and the required resources are ready.

\begin{algorithm}[!t]
\small
\KwData{$\{A_{i, j}\}$; $\{d_{i, j}\}$; $\{d_{i,j}^P\}$; $\{d_{i,j}^I\}$; $T$}
\KwResult{The optimal execution plan}
$\mathcal{S}(1) \gets \emptyset$ \;
\ForEach{model $i \in \mathcal{M}$}{
    \If{$d_{i, \, 1} \leq T$}{
    $\mathcal{S}(1) \gets \mathcal{S}(1) \cup \{(A_{i, 1}, ~ d^P_{i, 1}, ~ d_{i, 1}, ~ [i])\}$\;
    }
}
\For{$j=1$ to $r-1$}{
    $\mathcal{S}(j + 1) \gets \emptyset$ \;
    \ForEach{quaternion $(v, t^P, t, m\_list) \in \mathcal{S}(j)$}{
        \ForEach{model $i \in \mathcal{M}$}{
            $cur\_t \gets \max(t^p + d_{i, j+1},~t + d^I_{i, j+1})$ \;
            \If{$cur\_t \leq T$}{
                $cur\_v \gets v + A_{i, j+1}$ \;
                $cur\_t^P \gets t^P + d^P_{i, j+1} $ \;
                $m\_list.append(i)$ \;
                $\mathcal{S}(j+1) \gets \mathcal{S}(j+1) \cup \{(cur\_v, cur\_t^P, cur\_t, m\_list)\}$\;
            }
        }
    }
    Remove dominated execution plans from $\mathcal{S}(j+1)$ \;
}
Return the execution plan with the highest $v$ in $\mathcal{S}(r)$ \;
\caption{Dynamic Programming Algorithm}
\label{alg:dp}
\end{algorithm}

Let $d_{i,j}^P$ be the \emph{preprocessing delay} of analyzing SRoI $j$ with model $i$ and $d_{i,j}^I$ be its \emph{inference delay}. It follows that $\mathit{d_{i,\,j}} = d_{i,j}^P + d_{i,j}^I$. Assume that previous SRoIs $\{1, \cdots, j-1\}$ complete preprocessing at timestamp $t^P$ and complete inference at $t$. There are two cases depending on whether SRoI $j$ starts inference immediately after its preprocessing or waits. If there is no wait (e.g., SRoI $1$ and SRoI $3$ in Fig.~\ref{fig:pipeline}), the pipelined analysis latency for SRoIs $\{1, \cdots, j\}$ will be $t^P + \mathit{d_{i,\,j}}$; otherwise (e.g., SRoI $2$ and SRoI $4$ in Fig.~\ref{fig:pipeline}), the latency will be $t + d^I_{i,j}$. By integrating these two cases, the pipelined analysis latency can be updated as $\max(t^P + \mathit{d_{i,\,j}}, t + d^I_{i,j})$. Following this way, we can estimate the pipelined analysis latency $\mathcal{L(X)}$ of an execution plan $\mathcal{X}$.

We further design an efficient \emph{dynamic programming} algorithm to find the optimal execution plan for a set of SRoIs with a given processing order $\{1, 2, \cdots, r\}$. This algorithm is based on the concept of ``dominated pairs'' \cite{lawler1979fast}. Let $\mathcal{X}_j$ denote a \emph{feasible} execution plan for SRoIs $\{1, 2, \cdots, j\}$ that satisfies $\mathcal{L}(\mathcal{X}_j) \leq T$. For each feasible execution plan, we use a quaternion $(v, t^p, t, m\_list)$ to record its cumulative accuracy, preprocessing completion time, processing completion time, and allocated model list. We say a feasible execution plan $\mathcal{X}_j$ dominates another feasible execution plan $\mathcal{X}'_j$ if the following constraints are satisfied.
\begin{equation}
v(\mathcal{X}_j) \geq v(\mathcal{X}'_j), \quad
t^p(\mathcal{X}_j) \leq t^p(\mathcal{X}'_j), \quad
t(\mathcal{X}_j) \leq t(\mathcal{X}'_j)
\end{equation}

According to this definition, the dominated execution plans can be safely pruned. The details are shown in Algorithm~\ref{alg:dp}, where $\mathcal{S}(j)$ is the set of all feasible execution plans for SRoIs $\{1, 2, \cdots, j\}$. Note that this algorithm reports the optimal execution plan for a given processing order of SRoIs. Applying this algorithm to all possible processing orders can obtain the global optimal execution plan. Nevertheless, doing so incurs significant latency with only minor accuracy gains compared with directly using the execution plan reported on a randomly generated processing order. Thus, we finally choose the latter method to approximate the global optimal execution plan.


Based on the obtained execution plan, the \textbf{inference scheduler} acquires PIs and sends them to appropriate locations for inference. It then converts the inference results to SphBBs. Spherical NMS with a default threshold $0.6$ is further applied to the integrated SphBBs to prevent the same objects from being detected repeatedly.

\section{Performance Evaluation}
\subsection{System Implementation}

We implement a prototype of \texttt{OmniSense} with commodity devices. The mobile device is an Nvidia Jetson TX$2$ \cite{Jetson} mobile development board, a typical embedded smart computing device. The edge server is a desktop computer with an Intel i$7$-$6850$K CPU and an Nvidia GeForce GTX $1080$Ti GPU. The mobile device and the edge server are connected via an ASUS AC$1900$ router, and they both run the Ubuntu $18.04$ OS. The prototype is implemented in \texttt{Python} to allow easy integration with deep learning-based vision models. All video and image operations are implemented with the \texttt{OpenCV} library \cite{opencv}. Images, control messages, and detection results are passed between the mobile device and the edge server with a high-speed universal messaging library, \texttt{ZeroMQ}\cite{zeroMQ}.

\subsection{Evaluation Setup}

\noindent \textbf{Videos and models:} The video dataset and object detection models for evaluation are those demonstrated in TABLE~\ref{tab:video-dataset} and TABLE~\ref{tab:models}, respectively. Videos are stored on the mobile device in the ERP format and fed into the system frame by frame.

\noindent \textbf{Networks:} We employ the Linux traffic shaping tool \texttt{tc} to set the outgoing bandwidth of the mobile device to $17.9$ Mbps to match the average $5$G upload bandwidth of a major US mobile network provider (T-Mobile) \cite{upload-5g}. This value is used by default unless we study the influences of network settings.

\noindent \textbf{Performance Metrics:} ($1$) \emph{Spherical mAP (Sph-mAP)}. It is a spherical version of the standard mean Average Precision (mAP) metric \cite{everingham2015pascal} that is widely used for $2$D object detection. When applied to $360^\circ$ videos, the rectangular BB (regular IoU) is replaced with the SphBB (SphIoU) \cite{zhao2020spherical}. The Sph-mAP is calculated against the ground-truth detection results introduced in \cref{sec:motivation-setup}. ($2$) \emph{Mean End-to-end (E$2$E) latency}. It is the mean time taken to detect one frame, from feeding a frame into the system to obtaining the final detection results.

\noindent \textbf{Baselines:} ($1$) \emph{ERP}: This baseline directly feeds one ERP frame into an object detection model. To make the final results comparable, we further convert the detected rectangular BBs to SphBBs. ($2$) \emph{CubeMap:} It is based on the Cubemap projection \cite{cubemap-facebook, cheng2018cube} of $360^\circ$ videos. To be specific, it first projects the input frame onto a cube's six faces, and each face corresponds to a PI with a $90^\circ \times 90^\circ$ FoV. The PIs are subsequently analyzed by an object detection model. The detection results are further integrated and back-projected to the sphere to obtain the final detection results.

\noindent \textbf{Latency control:} To accommodate a wide range of latency budgets for various applications, \texttt{OmniSense} exposes an API to allow latency control. It is the per-frame analysis latency budget $T$ described in \cref{sec:model-allocation}. We investigate the performance of \texttt{OmniSense} with $T$ in a reasonable range from $500$ ms to $4,500$ ms. Specifically, we set the value of $T$ based on the end-to-end latencies of baselines to make the results comparable. Let $T_{ei}$ ($T_{ci}$) denote the $95\%$ of the mean E$2$E latency of ERP (CubeMap) using model $i$ for inference. We accordingly report the performance of \texttt{OmniSense} under the representative latency budgets $T_{e4}$, $T_{c2}$, $T_{c3}$, and $T_{c4}$.

\begin{figure}[!t]
    \centering
    \begin{subfigure}{.48\linewidth}
        \centering
        \includegraphics[width=\linewidth]{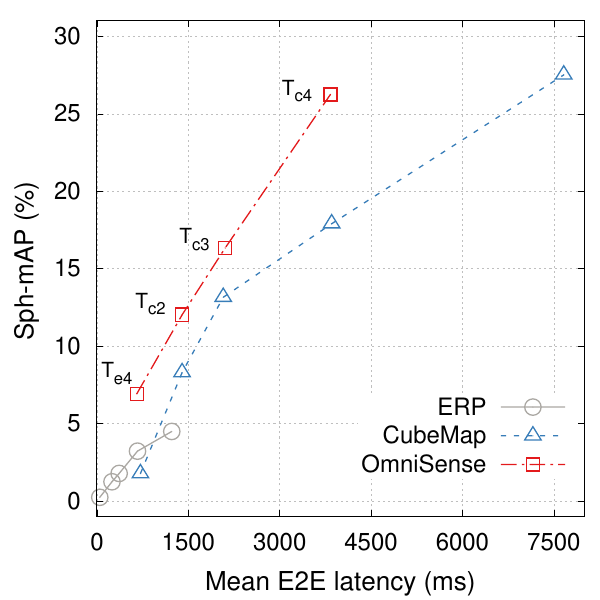}
        \caption{Video: New-Orleans-drive}
        \label{fig:orleans-res}
    \end{subfigure}
    \begin{subfigure}{.48\linewidth}
        \centering
        \includegraphics[width=\linewidth]{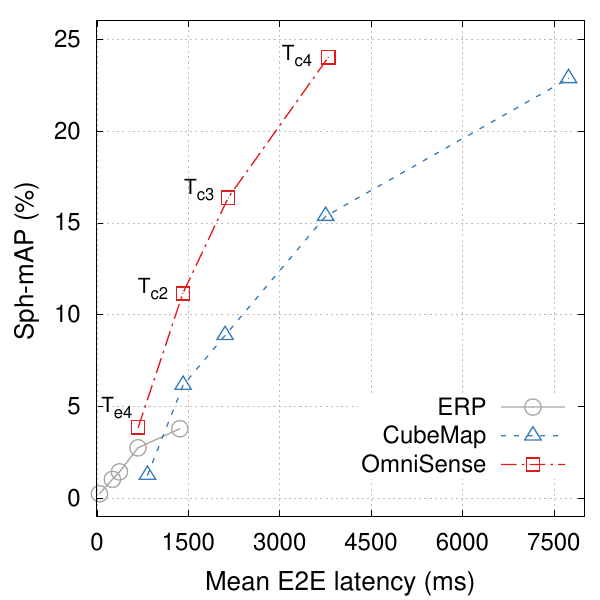}
        \caption{Video: Chicago-drive}
        \label{fig:chicago-res}
    \end{subfigure}
    \begin{subfigure}{.48\linewidth}
        \centering
        \includegraphics[width=\linewidth]{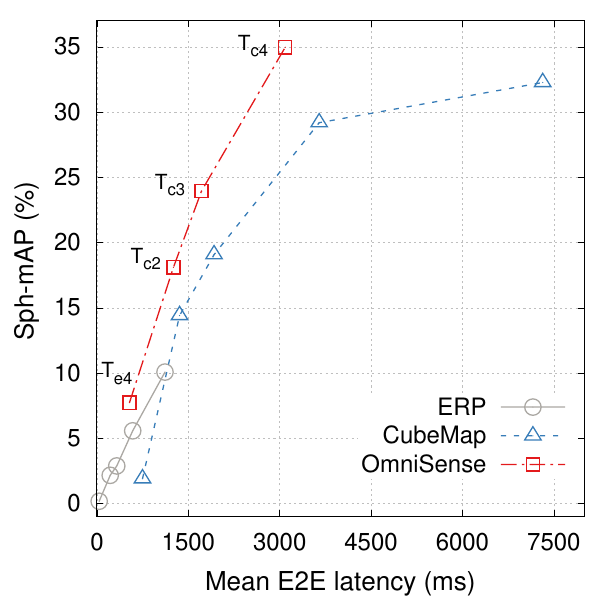}
        \caption{Video: Expressway-drive}
        \label{fig:express-res}
    \end{subfigure}
    \begin{subfigure}{.48\linewidth}
        \centering
        \includegraphics[width=\linewidth]{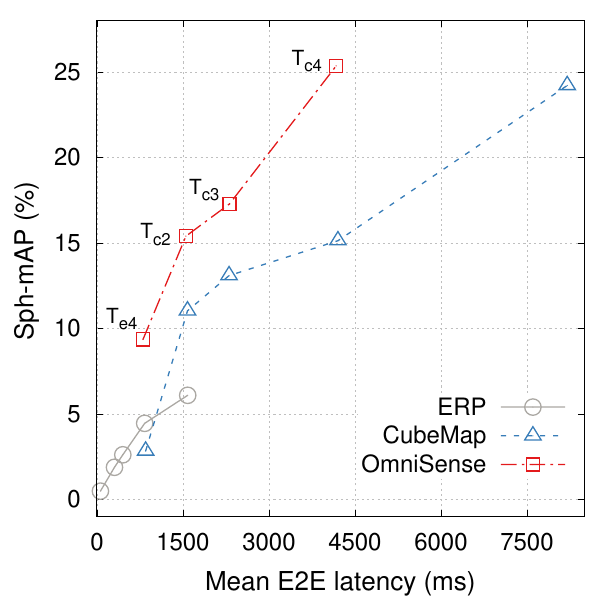}
        \caption{Video: Sunny-walk1}
        \label{fig:sunny-walk-res}
    \end{subfigure}
    \begin{subfigure}{.48\linewidth}
        \centering
        \includegraphics[width=\linewidth]{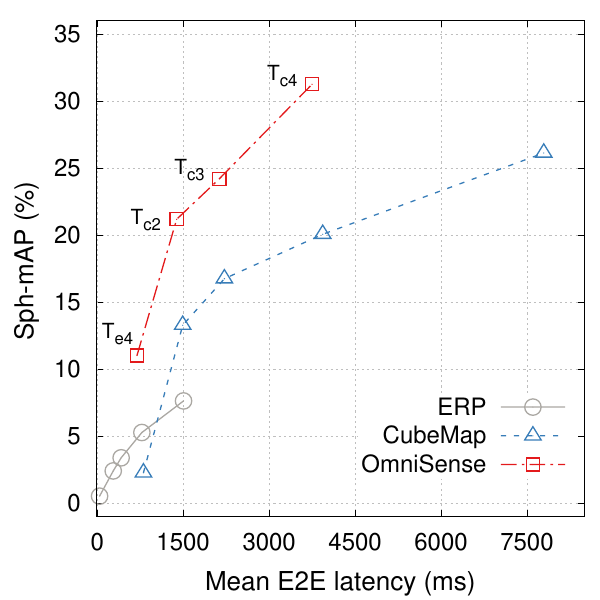}
        \caption{Video: Sunny-walk2}
        \label{fig:sunny-walk2-res}
    \end{subfigure}
    \begin{subfigure}{.48\linewidth}
        \centering
        \includegraphics[width=\linewidth]{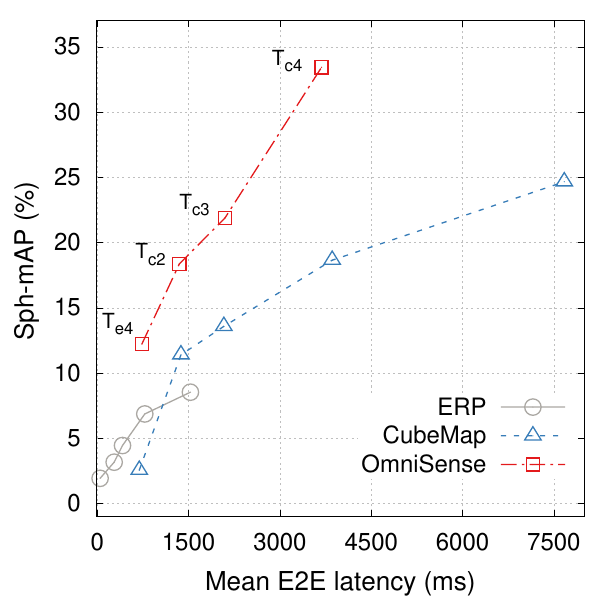}
        \caption{Video: Cloudy-walk}
        \label{fig:cloudy-walk-res}
    \end{subfigure}
    \caption{Overall performance comparisons of various methods.}
\label{fig:accuracy-latency}
\vspace{-0.5cm}
\end{figure}

\subsection{Evaluation Results}
\subsubsection{Performance Improvement}

We first present the overall performance of \texttt{OmniSense} as well as baselines on various videos in Fig.~\ref{fig:accuracy-latency}. As one can see, compared to the baseline with a similar mean E$2$E latency, \texttt{OmniSense} always yields higher accuracies, achieving $19.8\%$ -- $114.6\%$ accuracy improvements relatively, about $58.3\%$ on average. For example, with $T$ set to $T_{c2}$, \texttt{OmniSense} reaches a Sph-mAP of $11.2\%$ with a mean E$2$E latency of $1,409$ ms on the \texttt{Chicago-drive} video, while the corresponding CubeMap baseline only achieves a Sph-mAP of $6.2\%$ at the cost of $1,415$ ms. This indicates that \texttt{OmniSense} indeed successfully identified the ever-changing SRoIs and matched them with appropriate models. Meanwhile, benefiting from the more reasonable resource allocation strategy, \texttt{OminiSense} achieves an accuracy above or competitive with the highest achievable accuracy of baselines while significantly reducing the mean E$2$E latency ($2.0 \times$ -- $2.4 \times$ speedups). For example, it takes the CubeMap baseline $8,202$ ms to achieve a Sph-mAP of $24.2\%$ on the \texttt{Sunny-walk1} video, while \texttt{OmniSense} obtains a similar accuracy of $25.4\%$ with an approximate $2.0\times$ speedup ($4,173$ ms).

\subsubsection{System overhead on mobile devices}
\label{sec:sys-overhead}

\begin{figure}[!t]
    \centering
    \begin{subfigure}{.24\textwidth}
        \centering
        \includegraphics[width=\linewidth]{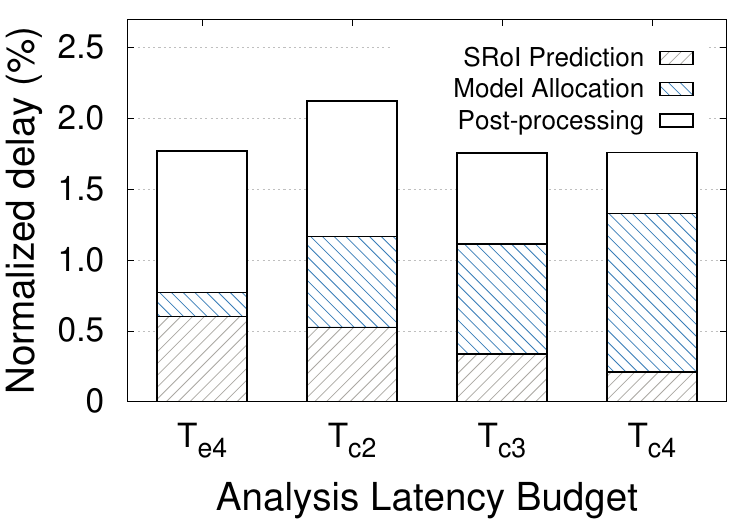}
        \caption{Video: Chicago-drive}
        \label{fig:drive-overhead}
    \end{subfigure}
    \begin{subfigure}{.24\textwidth}
        \centering
        \includegraphics[width=\linewidth]{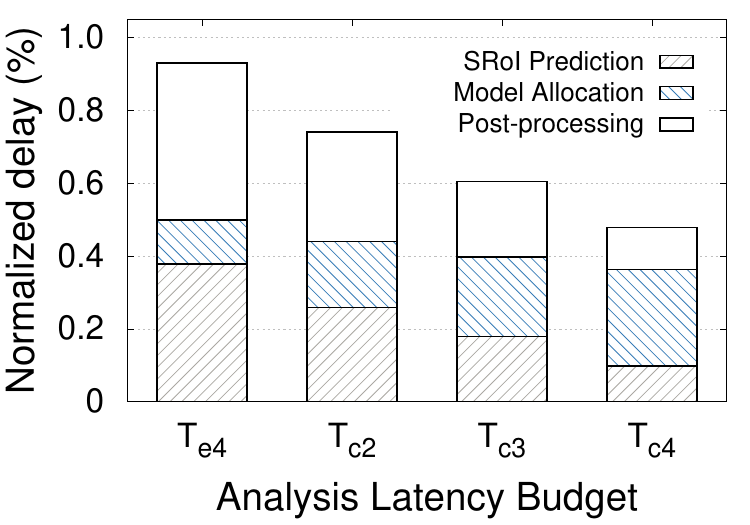}
        \caption{Video: Sunny-walk2}
        \label{fig:walk-overhead}
    \end{subfigure}
\caption{System overheads on the mobile device. The delays are normalized by the corresponding mean E$2$E latencies.}
\label{fig:sys-overhead}
\vspace{-0.4cm}
\end{figure}

The primary system overheads of \texttt{OmniSense} come from its SRoI prediction, model allocation, and post-processing. We demonstrate the breakdowns of their time consumption in Fig.~\ref{fig:sys-overhead}. As shown, the overall system overhead is within $2.5\%$ of the mean E$2$E latency for the \texttt{Chicago-drive} video and within $1\%$ for the \texttt{Sunny-walk2} video. With the same analysis latency budget setting, these two videos have comparable mean E$2$E latencies. Their overhead discrepancy is because the \texttt{Chicago-drive} video has more predicted SRoIs and SphBBs. Hence, it takes longer for the video to allocate vision models and transform the detected BBs, leading to higher system overheads. Despite this, the system overheads for both videos only account for a small fraction of the E$2$E latency. This means that the system design of \texttt{OmniSense} is lightweight and efficient and verifies the effectiveness of latency control.

\subsubsection{Sensitivity to image compression quality}

\begin{figure}[!t]
    \centering
    \begin{subfigure}{.25\textwidth}
        \centering
        \includegraphics[width=\linewidth]{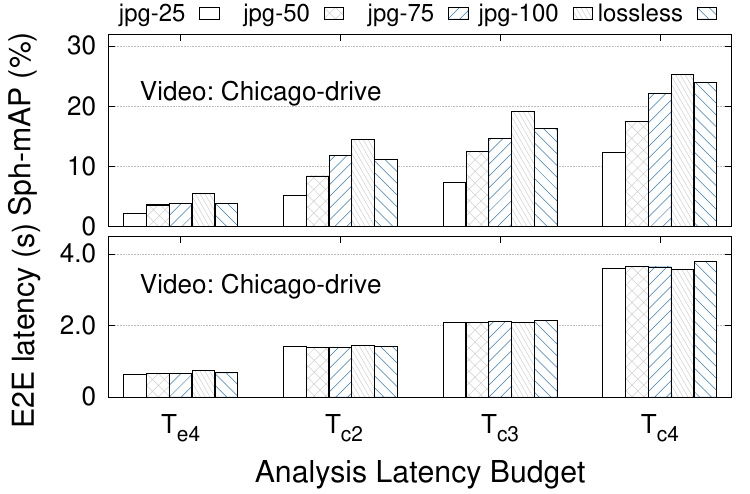}
        \caption{PI compression quality}
        \label{fig:drive-image-quality}
    \end{subfigure}
    \begin{subfigure}{.23\textwidth}
        \centering
        \includegraphics[width=\linewidth]{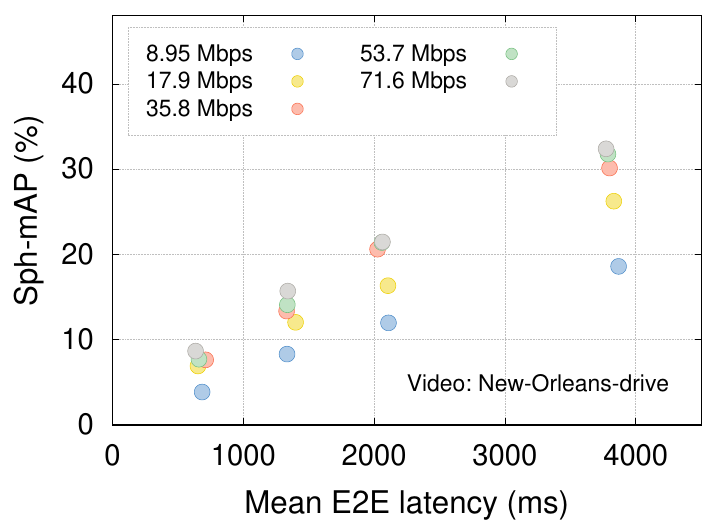}
        \caption{Network bandwidth}
        \label{fig:drive-bandwidth}
    \end{subfigure}
\caption{Sensitivities to image qualities and network conditions.}
\label{fig:sensitivity}
\vspace{-0.4cm}
\end{figure}

\texttt{OmniSense} compresses the projected PIs in the lossless PNG format by default before sending them to the edge server. Nonetheless, employing lossy image formats like JPEG can significantly decrease the image size, further reducing network delivery delays. The saved latency may be used for model upgrades, i.e., selecting a more accurate model with a higher latency, thus improving the overall accuracy. To study the impacts of PI compression quality, we adjust the JPEG compression quality parameter (higher values indicate higher image quality) and show the performance variations of \texttt{OmniSense} in Fig.~\ref{fig:drive-image-quality}.

As shown, \texttt{OmniSense} maintains a similar mean E$2$E latency under the same analysis latency budget regardless of the image compression quality. This indicates that \texttt{OmniSense} can adaptively use the saved latency to upgrade the analysis models. By trading image quality for more accurate models, moderate compression (i.e., jpg-$100$) can lead to an accuracy gain compared with lossless compression. However, aggressively compressing PIs does not always benefit accuracy. For instance, as the quality parameter decreases from $100$ to $25$, accuracy shows a downward trend and is consistently lower than that of lossless compression. This suggests that the accuracy drops caused by the severe image quality degradation can hardly be offset by the benefits brought by the reduced network delivery delays. Choosing a suitable PI compression quality to maximize accuracy requires careful trade-offs between the video content, network conditions, and model characteristics, and we leave it to future work.

\subsubsection{Sensitivity to network settings}
Fig.~\ref{fig:drive-bandwidth} demonstrates the impacts of varying network bandwidths on the performance of \texttt{OmniSense}. When the available bandwidth becomes scarce (i.e., $8.95$ Mbps), the network delivery delays increase. \texttt{OmniSense} adaptively trades accuracy for latency by switching to cheaper models. Conversely, when the available bandwidth becomes abundant (i.e., $35.8$ Mbps), the network delivery delays decrease. \texttt{OmniSense} takes advantage of the saved latency to improve accuracy by aggressively choosing more expensive models. This confirms that \texttt{OmniSense} can adapt to variations in network resources while seizing every chance to improve accuracy. Also, as the bandwidth increases (e.g., $> 35.8$ Mbps), the network delivery delay will no longer be the system bottleneck. Allocating excessive bandwidth in this case only results in marginal accuracy improvements.

\section{Conclusion}
Immersive video analytics will be essential in unlocking the potential of increasingly deployed omnidirectional cameras if the significant computation and network resource challenges can be addressed. Motivated by our measurement insights into diverse $360^\circ$ videos, we have proposed an edge-assisted immersive video analytics framework \texttt{OmniSense}. To adapt to the dynamic video content and network conditions, \texttt{OmniSense} introduces a lightweight algorithm to identify the ever-changing SRoIs. It then smartly allocates the most suitable local or remote model for each predicted SRoI to achieve low latency and high accuracy. Extensive evaluations have verified the effectiveness and superiority of \texttt{OmniSense}.

\clearpage
\bibliographystyle{IEEEtran}
\balance
\bibliography{reference}

\begin{thebibliography}{10}
\providecommand{\url}[1]{#1}
\csname url@samestyle\endcsname
\providecommand{\newblock}{\relax}
\providecommand{\bibinfo}[2]{#2}
\providecommand{\BIBentrySTDinterwordspacing}{\spaceskip=0pt\relax}
\providecommand{\BIBentryALTinterwordstretchfactor}{4}
\providecommand{\BIBentryALTinterwordspacing}{\spaceskip=\fontdimen2\font plus
\BIBentryALTinterwordstretchfactor\fontdimen3\font minus
  \fontdimen4\font\relax}
\providecommand{\BIBforeignlanguage}[2]{{%
\expandafter\ifx\csname l@#1\endcsname\relax
\typeout{** WARNING: IEEEtran.bst: No hyphenation pattern has been}%
\typeout{** loaded for the language `#1'. Using the pattern for}%
\typeout{** the default language instead.}%
\else
\language=\csname l@#1\endcsname
\fi
#2}}
\providecommand{\BIBdecl}{\relax}
\BIBdecl

\bibitem{insta360-one}
Insta360, ``Insta360 one x2,''
  \url{https://www.insta360.com/product/insta360-onex2}, [Online; accessed
  13-Jan-2023].

\bibitem{gopro-max}
``Gopro max 360 action camera,''
  \url{https://gopro.com/en/ca/shop/cameras/max/CHDHZ-202-master.html},
  [Online; accessed 13-Jan-2023].

\bibitem{qian2018flare}
F.~Qian, B.~Han, Q.~Xiao, and V.~Gopalakrishnan, ``Flare: Practical
  viewport-adaptive 360-degree video streaming for mobile devices,'' in
  \emph{Proceedings of the 24th Annual International Conference on Mobile
  Computing and Networking (MobiCom'18)}, 2018, pp. 99--114.

\bibitem{chen2021popularity}
X.~Chen, T.~Tan, and G.~Cao, ``Popularity-aware 360-degree video streaming,''
  in \emph{Proceedings of the IEEE Conference on Computer Communications
  (INFOCOM'21)}, 2021, pp. 1--10.

\bibitem{zhang2017live}
H.~Zhang, G.~Ananthanarayanan, P.~Bodik, M.~Philipose, P.~Bahl, and M.~J.
  Freedman, ``Live video analytics at scale with approximation and
  delay-tolerance,'' in \emph{Proceeding of the 14th USENIX Symposium on
  Networked Systems Design and Implementation (NSDI'17)}, 2017.

\bibitem{jiang2018chameleon}
J.~Jiang, G.~Ananthanarayanan, P.~Bodik, S.~Sen, and I.~Stoica, ``Chameleon:
  Scalable adaptation of video analytics,'' in \emph{Proceedings of the 2018
  Conference of the ACM Special Interest Group on Data Communication
  (SIGCOMM'18)}, 2018, pp. 253--266.

\bibitem{ran2018deepdecision}
X.~Ran, H.~Chen, X.~Zhu, Z.~Liu, and J.~Chen, ``Deepdecision: A mobile deep
  learning framework for edge video analytics,'' in \emph{Proceedings of
  INFOCOM}, 2018.

\bibitem{li2020reducto}
Y.~Li, A.~Padmanabhan, P.~Zhao, Y.~Wang, G.~H. Xu, and R.~Netravali, ``Reducto:
  On-camera filtering for resource-efficient real-time video analytics,'' in
  \emph{Proceedings of the Annual conference of the ACM Special Interest Group
  on Data Communication on the applications, technologies, architectures, and
  protocols for computer communication (SIGCOMM'20)}, 2020, pp. 359--376.

\bibitem{zhang2021elf}
W.~Zhang, Z.~He, L.~Liu, Z.~Jia, Y.~Liu, M.~Gruteser, D.~Raychaudhuri, and
  Y.~Zhang, ``Elf: Accelerate high-resolution mobile deep vision with
  content-aware parallel offloading,'' in \emph{Proceedings of the 27th Annual
  International Conference on Mobile Computing and Networking (MobiCom'21)},
  2021, pp. 201--214.

\bibitem{zhang2018awstream}
B.~Zhang, X.~Jin, S.~Ratnasamy, J.~Wawrzynek, and E.~A. Lee, ``Awstream:
  Adaptive wide-area streaming analytics,'' in \emph{Proceedings of the 2018
  Conference of the ACM Special Interest Group on Data Communication
  (SIGCOMM'18)}, 2018, pp. 236--252.

\bibitem{guan2019pano}
Y.~Guan, C.~Zheng, X.~Zhang, Z.~Guo, and J.~Jiang, ``Pano: Optimizing
  360$^\circ$ video streaming with a better understanding of quality
  perception,'' in \emph{Proceedings of the ACM Special Interest Group on Data
  Communication (SIGCOMM'19)}, 2019, pp. 394--407.

\bibitem{hu2017deep}
H.-N. Hu, Y.-C. Lin, M.-Y. Liu, H.-T. Cheng, Y.-J. Chang, and M.~Sun, ``Deep
  360 pilot: Learning a deep agent for piloting through 360$^{\circ}$ sports
  videos,'' in \emph{Proceedings of the IEEE Conference on Computer Vision and
  Pattern Recognition (CVPR'17)}, 2017, pp. 1396--1405.

\bibitem{cubemap-facebook}
Facebook, ``Under the hood: Building 360 video,''
  \url{https://code.facebook.com/posts/1638767863078802/under-the-hood-building-360-video/},
  2015, [Online; accessed 13-Jan-2023].

\bibitem{sun2017learning}
Y.-C. Sun and K.~Grauman, ``Learning spherical convolution for fast features
  from 360$^{\circ}$ imagery,'' in \emph{Proceedings of the 31st Conference on
  Neural Information Processing Systems (NeurIPS'17)}, 2017, pp. 529--539.

\bibitem{coors2018spherenet}
B.~Coors, A.~P. Condurache, and A.~Geiger, ``Spherenet: Learning spherical
  representations for detection and classification in omnidirectional images,''
  in \emph{Proceedings of the European Conference on Computer Vision
  (ECCV'18)}, 2018, pp. 518--533.

\bibitem{chou2020360}
S.-H. Chou, C.~Sun, W.-Y. Chang, W.-T. Hsu, M.~Sun, and J.~Fu, ``360-indoor:
  Towards learning real-world objects in 360$^{\circ}$ indoor equirectangular
  images,'' in \emph{Proceedings of the IEEE/CVF Winter Conference on
  Applications of Computer Vision (WACV'20)}, 2020, pp. 845--853.

\bibitem{lee2019spherephd}
Y.~Lee, J.~Jeong, J.~Yun, W.~Cho, and K.-J. Yoon, ``Spherephd: Applying cnns on
  a spherical polyhedron representation of 360$^{\circ}$ images,'' in
  \emph{Proceedings of the IEEE/CVF Conference on Computer Vision and Pattern
  Recognition (CVPR'19)}, 2019, pp. 9181--9189.

\bibitem{yang2018object}
W.~Yang, Y.~Qian, J.-K. K{\"a}m{\"a}r{\"a}inen, F.~Cricri, and L.~Fan, ``Object
  detection in equirectangular panorama,'' in \emph{Proceedings of the 24th
  International Conference on Pattern Recognition}, 2018, pp. 2190--2195.

\bibitem{eder2020tangent}
M.~Eder, M.~Shvets, J.~Lim, and J.-M. Frahm, ``Tangent images for mitigating
  spherical distortion,'' in \emph{Proceedings of the IEEE/CVF Conference on
  Computer Vision and Pattern Recognition (CVPR'20)}, 2020, pp.
  12\,426--12\,434.

\bibitem{ananthanarayanan2017real}
G.~Ananthanarayanan, P.~Bahl, P.~Bod{\'\i}k, K.~Chintalapudi, M.~Philipose,
  L.~Ravindranath, and S.~Sinha, ``Real-time video analytics: The killer app
  for edge computing,'' \emph{computer}, vol.~50, no.~10, pp. 58--67, 2017.

\bibitem{wang2019bridging}
Y.~Wang, W.~Wang, J.~Zhang, J.~Jiang, and K.~Chen, ``Bridging the edge-cloud
  barrier for real-time advanced vision analytics,'' in \emph{Proceedings of
  the 11th USENIX Workshop on Hot Topics in Cloud Computing (HotCloud'19)},
  2019.

\bibitem{guo2021crossroi}
H.~Guo, S.~Yao, Z.~Yang, Q.~Zhou, and K.~Nahrstedt, ``Crossroi: Cross-camera
  region of interest optimization for efficient real time video analytics at
  scale,'' in \emph{Proceedings of the 12th ACM Multimedia Systems Conference
  (MMSys'21)}, 2021, pp. 186--199.

\bibitem{jiang2018mainstream}
A.~H. Jiang, D.~L.-K. Wong, C.~Canel, L.~Tang, I.~Misra, M.~Kaminsky, M.~A.
  Kozuch, P.~Pillai, D.~G. Andersen, and G.~R. Ganger, ``Mainstream: Dynamic
  stem-sharing for multi-tenant video processing,'' in \emph{Proceedings of the
  2018 USENIX Annual Technical Conference (ATC'18)}, 2018, pp. 29--42.

\bibitem{jiang2021flexible}
S.~Jiang, Z.~Lin, Y.~Li, Y.~Shu, and Y.~Liu, ``Flexible high-resolution object
  detection on edge devices with tunable latency,'' in \emph{Proceedings of the
  27th Annual International Conference on Mobile Computing and Networking
  (MobiCom'21)}, 2021, pp. 559--572.

\bibitem{canel2019scaling}
C.~Canel, T.~Kim, G.~Zhou, C.~Li, H.~Lim, D.~G. Andersen, M.~Kaminsky, and
  S.~R. Dulloor, ``Scaling video analytics on constrained edge nodes,'' in
  \emph{Proceedings of the 2nd SysML Conference (SysML'19)}, 2019.

\bibitem{emmons2019cracking}
J.~Emmons, S.~Fouladi, G.~Ananthanarayanan, S.~Venkataraman, S.~Savarese, and
  K.~Winstein, ``Cracking open the dnn black-box: Video analytics with dnns
  across the camera-cloud boundary,'' in \emph{Proceedings of the 2019 Workshop
  on Hot Topics in Video Analytics and Intelligent Edges (HotEdgeVideo'19)},
  2019, pp. 27--32.

\bibitem{du2020server}
K.~Du, A.~Pervaiz, X.~Yuan, A.~Chowdhery, Q.~Zhang, H.~Hoffmann, and J.~Jiang,
  ``Server-driven video streaming for deep learning inference,'' in
  \emph{Proceedings of the Annual conference of the ACM Special Interest Group
  on Data Communication on the applications, technologies, architectures, and
  protocols for computer communication (SIGCOMM'20)}, 2020, pp. 557--570.

\bibitem{wang2021scaled}
C.-Y. Wang, A.~Bochkovskiy, and H.-Y.~M. Liao, ``Scaled-yolov4: Scaling cross
  stage partial network,'' in \emph{Proceedings of the IEEE/CVF Conference on
  Computer Vision and Pattern Recognition (CVPR'21)}, 2021, pp.
  13\,029--13\,038.

\bibitem{new-orleans-driving}
J.~Utah, ``New orleans 8k - vr 360$^\circ$ drive - 60fps,''
  \url{https://www.youtube.com/watch?v=bSV8qc2_qFs}, 2019, [Online; accessed
  13-Jan-2023].

\bibitem{expressway-driving}
ActionKid, ``360$^\circ$ driving elmhurst, queens to bay ridge, brooklyn via
  brooklyn queens expressway (april 5, 2020),''
  \url{https://www.youtube.com/watch?v=gwgm8XFEmYE}, 2020, [Online; accessed
  13-Jan-2023].

\bibitem{chicago-driving}
J.~Utah, ``Chicago 8k 360$^\circ$ video - virtual reality - driving downtown,''
  \url{https://www.youtube.com/watch?v=Gu1D3BnIYZg}, 2020, [Online; accessed
  13-Jan-2023].

\bibitem{lin2014microsoft}
T.-Y. Lin, M.~Maire, S.~Belongie, J.~Hays, P.~Perona, D.~Ramanan,
  P.~Doll{\'a}r, and C.~L. Zitnick, ``Microsoft coco: Common objects in
  context,'' in \emph{European conference on computer vision (ECCV'14)}.\hskip
  1em plus 0.5em minus 0.4em\relax Springer, 2014, pp. 740--755.

\bibitem{zhao2020spherical}
P.~Zhao, A.~You, Y.~Zhang, J.~Liu, K.~Bian, and Y.~Tong, ``Spherical criteria
  for fast and accurate 360$^{\circ}$ object detection,'' in \emph{Proceedings
  of the 34th AAAI Conference on Artificial Intelligence (AAAI'20)}, 2020, pp.
  12\,959--12\,966.

\bibitem{pisinger1995minimal}
D.~Pisinger, ``A minimal algorithm for the multiple-choice knapsack problem,''
  \emph{European Journal of Operational Research}, vol.~83, no.~2, pp.
  394--410, 1995.

\bibitem{lawler1979fast}
E.~L. Lawler, ``Fast approximation algorithms for knapsack problems,''
  \emph{Mathematics of Operations Research}, vol.~4, no.~4, pp. 339--356, 1979.

\bibitem{Jetson}
Nvidia, ``Jetson tx2 module,''
  \url{https://developer.nvidia.com/embedded/jetson-tx2}, [Online; accessed
  13-Jan-2023].

\bibitem{opencv}
OpenCV, ``Opencv: Open source computer vision library,''
  \url{https://opencv.org/}, [Online; accessed 13-Jan-2023].

\bibitem{zeroMQ}
ZeroMQ, ``An open-source universal messaging library,''
  \url{https://zeromq.org/}, [Online; accessed 13-Jan-2023].

\bibitem{upload-5g}
Opensignal, ``Usa 5g experience report, january 2022,''
  \url{https://www.opensignal.com/reports/2022/01/usa/mobile-network-experience-5g},
  [Online; accessed 13-Jan-2023].

\bibitem{everingham2015pascal}
M.~Everingham, S.~Eslami, L.~Van~G., C.~K. Williams, J.~Winn, and A.~Zisserman,
  ``The pascal visual object classes challenge: A retrospective,''
  \emph{International journal of computer vision}, vol. 111, no.~1, pp.
  98--136, 2015.

\bibitem{cheng2018cube}
H.-T. Cheng, C.-H. Chao, J.-D. Dong, H.-K. Wen, T.-L. Liu, and M.~Sun, ``Cube
  padding for weakly-supervised saliency prediction in 360$^{\circ}$ videos,''
  in \emph{Proceedings of the IEEE Conference on Computer Vision and Pattern
  Recognition (CVPR'18)}, 2018, pp. 1420--1429.

\end{thebibliography}

\end{document}